\documentclass[a4paper,12pt]{article}
\pdfoutput=1
\usepackage{graphicx, rotating}
\usepackage{hyperref}
\usepackage{slashed}
\usepackage{ifpdf}

\ifx\pdfoutput\undefined
\usepackage[dvips,bookmarks=false]{hyperref}	
\else
\usepackage{hyperref}	
\fi
\hypersetup{colorlinks,bookmarksopen,bookmarksnumbered,citecolor=verdes,
linkcolor=blus,pdfstartview=FitH,urlcolor=rossos}
\def\hhref#1{\href{http://arxiv.org/abs/#1}{#1}} 

\usepackage{amsmath}
\usepackage{slashed}

\newcommand{\beq}{\begin{equation}}
\newcommand{\eeq}{\end{equation}}
\newcommand{\fig}[1]{~\ref{fig:#1}}

\newcount\Mac  \Mac=1  
\newcommand{\ifMac}[2]{\ifnum\Mac=1 #1 \else #2 \fi}
\def\putps(#1,#2)(#3,#4)#5#6{\ifnum\Mac=1 \put(#1,#2){\special{picture #5}}
\else  \put(#3,#4){\includegraphics{#6}} \fi}

\newcommand{\One}{\hbox{1\kern-.24em I}}

\renewcommand{\Im}{\mathop{\rm Im}}

\newcommand{\TeV}{\,{\rm TeV}}

\newcommand{\PRL}{Phys. Rev. Lett.}
\newcommand{\PL}{Phys. Lett.}
\newcommand{\PR}{Phys. Rev.}

\newcommand{\eq}[1]{~{\rm (\ref{eq:#1})}}

\newcommand{\lascia}[1]{}
\makeatletter
\def\art{\@ifnextchar[{\eart}{\oart}}
\def\eart[#1]#2#3#4#5#6{{\rm #2}, {#3 #4} {\rm (#6) #5} [arXiv:{\hhref{#1}}]}
\def\hepart[#1]#2{{\rm #2, arXiv:\hhref{#1}}}
\newcommand{\oart}[5]{{\rm #1}, {#2 #3} {\rm (#5) #4}}

\newcounter{alphaequation}[equation]
\def\thealphaequation{\theequation\hbox to
0.6em{\hfil\alph{alphaequation}\hfil}}
\def\eqnsystem#1{
\def\@eqnnum{{\rm (\thealphaequation)}}
\def\@@eqncr{\let\@tempa\relax \ifcase\@eqcnt \def\@tempa{& & &} \or
  \def\@tempa{& &}\or \def\@tempa{&}\fi\@tempa
  \if@eqnsw\@eqnnum\refstepcounter{alphaequation}\fi
\global\@eqnswtrue\global\@eqcnt=0\cr}
\refstepcounter{equation} \let\@currentlabel\theequation \def\@tempb{#1}
\ifx\@tempb\empty\else\label{#1}\fi
\refstepcounter{alphaequation}
\let\@currentlabel\thealphaequation
\global\@eqnswtrue\global\@eqcnt=0 \tabskip\@centering\let\\=\@eqncr
$$\halign to \displaywidth\bgroup \@eqnsel\hskip\@centering
$\displaystyle\tabskip\z@{##}$&\global\@eqcnt\@ne
\hskip2\arraycolsep\hfil${##}$\hfil& \global\@eqcnt\tw@\hskip2\arraycolsep
$\displaystyle\tabskip\z@{##}$\hfil
\tabskip\@centering&\llap{##}\tabskip\z@\cr}

\def\endeqnsystem{\@@eqncr\egroup$$\global\@ignoretrue} \makeatother

\def\Lag{{\cal L}}

\def\circa#1{\,\raise.3ex\hbox{$#1$\kern-.75em\lower1ex\hbox{$\sim$}}\,}

\usepackage{multicol}
\usepackage{color}
\definecolor{rosso}{cmyk}{0,1,1,0.4}
\definecolor{rossos}{cmyk}{0,1,1,0.55}
\definecolor{rossoc}{cmyk}{0,1,1,0.2}
\definecolor{blu}{cmyk}{1,1,0,0.3}
\definecolor{blus}{cmyk}{1,1,0,0.6}
\definecolor{bluc}{cmyk}{1,1,0,0.1}
\definecolor{verde}{cmyk}{0.92,0,0.59,0.25}
\definecolor{verdec}{cmyk}{0.92,0,0.59,0.15}
\definecolor{verdes}{cmyk}{0.92,0,0.59,0.4}
\definecolor{grigio}{cmyk}{0,0,0,0.07}
\definecolor{rosa}{cmyk}{0,0.1,0.1,0.02}
\definecolor{rosino}{cmyk}{0,0.05,0.05,0.02}
\definecolor{rosas}{cmyk}{0,0.3,0.25,0.05}
\definecolor{celeste}{cmyk}{0.1,0,0,0.02}
\definecolor{giallino}{cmyk}{0,0,0.4,0.02}
\definecolor{rosso}{cmyk}{0,1,1,0.4}
\definecolor{rossos}{cmyk}{0,1,1,0.55}
\definecolor{rossoc}{cmyk}{0,1,1,0.2}
\definecolor{blu}{cmyk}{1,1,0,0.3}
\definecolor{bluc}{cmyk}{1,1,0,0.1}
\definecolor{blucc}{cmyk}{0.7,0.5,0,0}
\definecolor{viola}{cmyk}{0,1,0,0.6}
\definecolor{viola2}{cmyk}{0,1,0.2,0.6}
\definecolor{verde}{cmyk}{0.92,0,0.59,0.25}
\definecolor{verdec}{cmyk}{0.92,0,0.59,0.15}
\definecolor{verdes}{cmyk}{0.92,0,0.59,0.4}
\definecolor{verdino}{cmyk}{0.12,0,0.09,0.05}
\definecolor{giallo}{cmyk}{0,0,1,0}
\definecolor{gialloverde}{cmyk}{0.44,0,0.74,0}

\font\tenrsfs=rsfs10 at 12pt

\font\sevenrsfs=rsfs7
\font\fiversfs=rsfs5
\newfam\rsfsfam
\textfont\rsfsfam=\tenrsfs
\scriptfont\rsfsfam=\sevenrsfs
\scriptscriptfont\rsfsfam=\fiversfs
\def\mathscr#1{{\fam\rsfsfam\relax#1}}
\def\Lag{\mathscr{L}}

\begin{document}
 IFUP-TH/2010-43\hfill CERN-PH-TH/2010-286
\color{black}
\vspace{1cm}
\begin{center}
{\Huge\bf\color{black}LHC bounds on large extra dimensions}\\
\bigskip\color{black}\vspace{0.6cm}{
{\large\bf Roberto Franceschini$^a$, Pier Paolo Giardino$^b$}\\[2mm]
{\large\bf Gian F.\ Giudice$^{c}$, Paolo Lodone$^d$, Alessandro Strumia$^{b,e}$}
} \\[7mm]
{\it (a) Institut de Th\'eorie des Ph\'enom\`enes Physiques, EPFL,
CH-1015 Lausanne, Switzerland}\\[3mm]
{\it (b) Dipartimento di Fisica dell'Universit{\`a} di Pisa and INFN, Italia}\\[3mm]
{\it (c) CERN, Theory Division, CH-1211, Geneva 23, Switzerland}\\[3mm]
{\it (d) Scuola Normale Superiore di Pisa and INFN, Italia}\\[3mm]
{\it (e) NICPB, Ravala 10, 10143 Tallinn, Estonia}\\
\end{center}
\bigskip
\centerline{\large\bf\color{blus} Abstract}
\begin{quote}\large
We derive new dominant bounds
on  the coefficient of the effective operator generated by
tree-level graviton exchange in large extra dimensions
from $pp\to jj$ data at LHC:
$M_T>2.1\TeV$ (ATLAS after 3.1/pb of integrated luminosity), $M_T>3.4\TeV$ (CMS after 36/pb), $M_T>3.2\TeV$ (ATLAS after 36/pb).
We clarify the role of on-shell graviton exchange and compare the
full graviton amplitude to data, setting bounds on the fundamental
quantum-gravity scale.
\color{black}
\end{quote}

\newpage

\section{Introduction}

With the start of the LHC program, experiments are already testing directly some of the theoretical ideas about new physics at the electroweak scale. In one popular scenario, which will be considered in this paper, Standard Model fields are
confined on a $3$-dimensional brane, while gravity propagates in the full 
$D$-dimensional space, with
$\delta$ flat and compactified
extra spatial dimensions ($D=4+\delta$)~\cite{add}. 
This scenario allows for quantum gravity at the weak scale and could therefore be a solution to the Higgs
mass hierarchy problem.  
Even without knowledge of the exact model for quantum gravity at the weak scale, we can make some definite predictions for collider experiments using either low-energy effective theory or semi-classical approximation, which can provide valid descriptions in certain kinematical domains. When experimental data are compared with expectations, it is then important to assess the validity of the approximations used in the theoretical calculations.

We can identify five different kinds of LHC signals which allow for a theoretical interpretation in terms of $D$-dimensional gravity.
\begin{enumerate}
\item {\it Missing $p_T$ from emission of massive gravitons} constituting the Kaluza-Klein tower. This signal is within control of the low-energy effective theory as long as the graviton energy is less than an ultraviolet cutoff $\Lambda_{\rm eff}$, which characterizes the onset of the new quantum-gravity theory. Validity of the perturbative expansion sets an upper bound on the cutoff 
\beq
\Lambda_{\rm eff}< [\Gamma (2+\delta /2)]^{\frac{1}{2+\delta}} (4\pi)^{\frac{4+\delta}{4+2\delta}} M_D ,
\eeq
where $M_D$ is the $D$-dimensional Planck mass in the notation of~\cite{noi}. This upper bound is saturated only when gravitons become fully strongly-interacting before entering the new regime of the underlying theory, and thus $\Lambda_{\rm eff}$ could actually turn out to be much smaller. This does not mean that missing  $p_T$ signals above $\Lambda_{\rm eff}$ vanish, but simply that they are not calculable without knowledge of the full theory.

\item {\it Tree-level exchange of gravitons} (fig.\fig{treeloop}a) generating the effective dimension-8 
operator $\cal T$~\cite{noi,han,hewett} 
\beq
\Lag_{\rm int} = c_{\cal T} \times~{\cal T} = 
 \frac{8}{M_{T}^4} \times \frac{1}{2}\left( T_{\mu\nu}T^{\mu\nu} -\frac{T_\mu^\mu T_\nu^\nu}{\delta +2} 
 \right) ,
\label{eq:tau}
\eeq
where $T_{\mu \nu}$ is the SM energy-momentum tensor. As discussed in section~\ref{th}, in most cases the dominant contribution to this operator comes from the ultraviolet end of the graviton spectrum. Therefore the parameter $M_{T}$ cannot be computed without knowledge of the underlying quantum-gravity theory. The case $\delta =1$ (and, to a certain extent, $\delta =2$) provides an interesting exception.

\item {\it Virtual graviton exchanges at one-loop level} (fig.\fig{treeloop}b) can become more important than tree-level effects because they induce dimension-6 effective operators, as opposed to the dimension-8 $\cal T$ operator~\cite{stru}. For pure graviton virtual intermediate states, a unique dimension-6 operator is generated    
\beq\label{eq:Upsilon}
\Lag =c_\Upsilon\times  \Upsilon ,~~~~\Upsilon =\frac{1}{2}
\bigg( \sum_f {\bar f}\gamma_\mu \gamma_5 f\bigg)
\bigg( \sum_f {\bar f}\gamma^\mu \gamma_5 f\bigg) ,
\eeq
where $f$ is any SM quark or lepton.
As in the case of tree-level graviton exchange, the coefficient $c_\Upsilon$ is fully sensitive to the ultraviolet completion of the theory and can be related to the fundamental parameters $M_D$ and $\delta$ only by specifying a cutoff procedure.

\item {\it Dijet events at large invariant mass and large rapidity separation}. In this kinematic regime, gravitational scattering can be reliably computed in the eikonal approximation~\cite{trans}. This is because scattering processes at center-of-mass energy larger than $M_D$ (the so-called transplanckian region) are governed by classical dynamics and any quantum-gravity effect is subdominant.

\begin{figure}[t]
$$\includegraphics{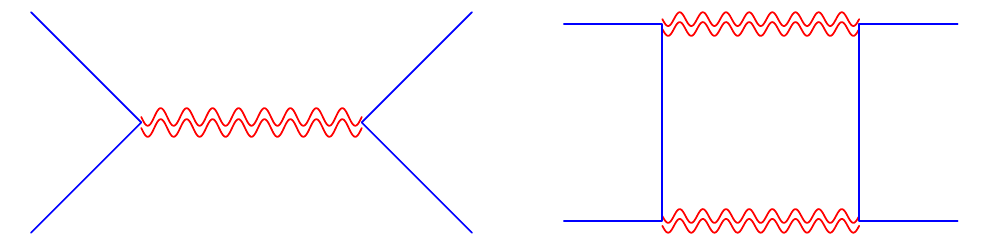}$$
\caption{\em Fig.~\ref{fig:treeloop}{\rm a}:
Tree-level graviton exchange generating the dimension-8 operator $\cal T$.
Fig.~\ref{fig:treeloop}{\rm b}: One-loop graviton exchange generating 
the dimension-6 operator $\Upsilon$.
\label{fig:treeloop}}
\end{figure}

\begin{table}[p]
\centering
\begin{tabular}{|c|c|cc|}
\hline\hline
Experiment&Process & + & $-$\\ 
\hline
LEP~\cite{lepc} & $e^+e^-\to \gamma \gamma$ & 0.93\TeV&1.01\TeV\\
LEP~\cite{lepcc} & $e^+e^-\to  e^+e^-$ & 1.18\TeV & 1.17\TeV\\
H1~\cite{h1} & $e^+p$ and $e^-p$ & 0.74\TeV & 0.71\TeV\\
ZEUS~\cite{zeus} & $e^+p$ and $e^-p$ & 0.72\TeV & 0.73\TeV\\
CDF~\cite{landsberg} &  $p\bar p\to e^+e^-,\gamma \gamma$& 0.99\TeV & 0.96\TeV\\ 
D\O~\cite{landsberg} & $p\bar p\to e^+e^-,\gamma \gamma$ & 1.28\TeV & 1.14\TeV\\
D\O~\cite{D0jj} &  $p\bar p\to jj$& 1.48 \TeV& 1.48\TeV\\ 
CMS at 7 TeV with 40/pb~\cite{CMSmu} & $pp\to \mu^-\mu^+$ &1.6\TeV&1.6\TeV\\
CMS at 7 TeV with 36/pb~\cite{CMSgammagamma} & $pp\to \gamma\gamma$ &1.74\TeV&1.71\TeV\\
\color{red}ATLAS at 7 TeV with 3.1/pb & \color{red} $pp\to jj$ & \color{red}2.2\TeV&\color{red}2.1\TeV\\
\color{red}ATLAS at 7 TeV with 36/pb & \color{red} $pp\to jj$ & \color{red}4.2\TeV&\color{red}3.2\TeV\\
\color{red}CMS at 7 TeV with 36/pb & \color{red} $pp\to jj$ & \color{red}4.2\TeV&\color{red}3.4\TeV\\
\hline
\end{tabular}
$$\includegraphics[width=0.7\textwidth]{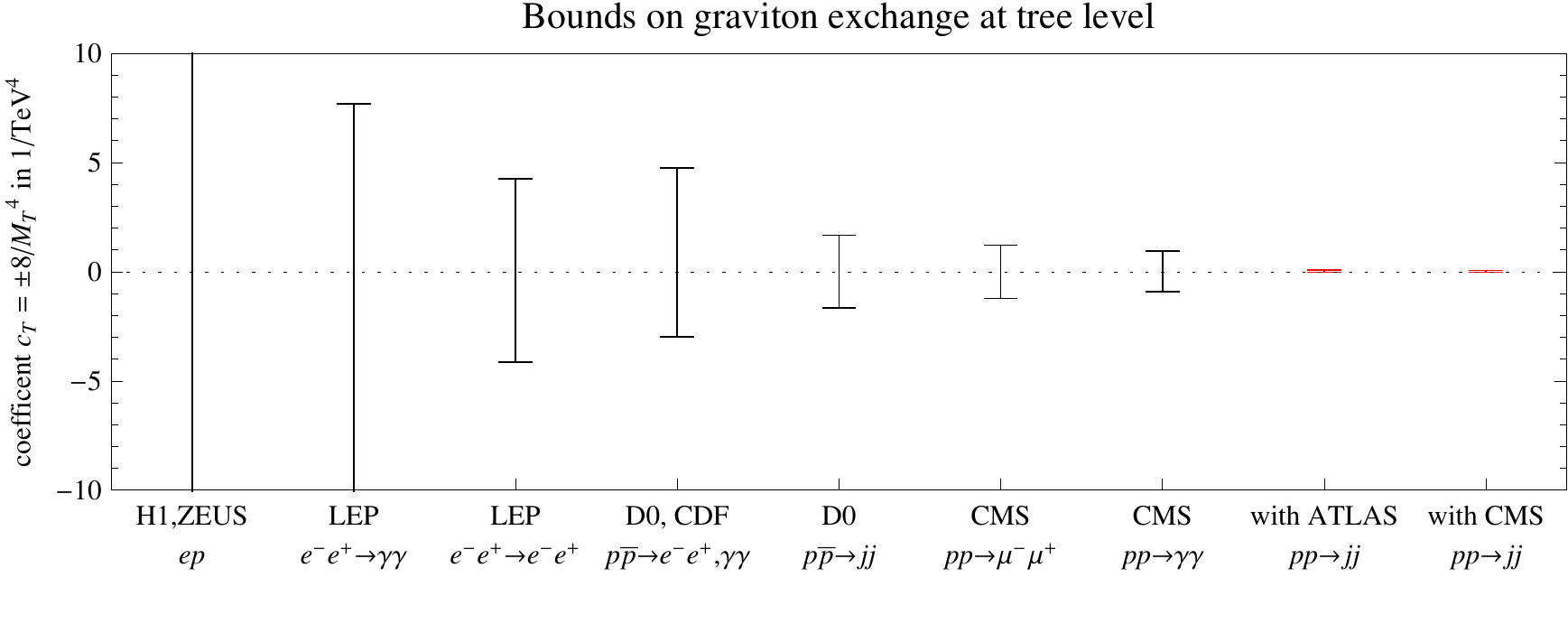}$$
\caption{\label{tab:tau} {\bf Tree-level graviton exchange}:
\em {95\% CL limits on the coefficient $M_{T}$ (known as Hewett normalization~\cite{hewett}) of the
dimension-8 operator $\cal T$ of eq.\eq{tau} for positive
and negative interference. The last two limits are derived in this work.
}}
\vspace{1cm}
\begin{tabular}{|c|c|cc|}
\hline\hline
Experiment&Process & + & $-$\\ 
\hline
LEP combined~\cite{lepcomb} & $e^+e^-\to e^+e^-$ & 11.3 & 11.5 \\
LEP combined~\cite{lepcomb} & $e^+e^-\to \mu^+\mu^-$ & 16.4 & 12.7 \\
LEP combined~\cite{lepcomb} & $e^+e^-\to \ell^+\ell^-$ & 17.2 & 15.1 \\
LEP combined~\cite{lepcomb} & $e^+e^-\to b\bar b$ & 15.3 & 11.5 \\
H1~\cite{h1} & $e^+p$ and $e^-p$ & 2.5 & 3.9\\
ZEUS~\cite{zeus} & $e^+p$ and $e^-p$ & 4.6 & 5.3\\
D\O~\cite{talk} &  $p\bar p\to e^+e^-$  & 4.7 & 5.5 \\
CDF~\cite{talk} & $p\bar p\to\ell^+\ell^-$ & 4.5 & 5.6 \\
CCFR~\cite{ccfr} & $\nu N$ scattering & 3.7 & 5.9 \\
D\O~\cite{talk} & $p\bar p\to jj$ & 3.2 & 3.1 \\
\color{red}ATLAS at 7 TeV with 3.1/pb  & \color{red}$pp\to jj$ & \color{red}5.3 & \color{red}4.2\\
\color{red}CMS at 7 TeV with 36/pb & \color{red} $pp\to jj$ & \color{red}11&\color{red}8.1\\
\hline
\multicolumn{2}{|c|}{combined}&22.4&15.7\\
\hline\hline
\end{tabular}
\caption{{\bf Loop-level graviton exchange}:
\em { 95\% CL limits on the coefficient 
$|c_\Upsilon/4\pi|^{-1/2}$ (in {\rm TeV}) of the dimension-6 operator $\Upsilon$ of eq.\eq{Upsilon}
for positive and negative values of $c_\Upsilon$.
\label{tab:dim6}}
}
%
\end{table}

\item {\it Black holes}. Black-hole formation and decay is expected to occur in the transplanckian region when the impact parameter becomes smaller than the corresponding Schwarzschild radius~\cite{DL}. Therefore it supplants gravitational scattering, in the limit of small rapidity separation. While transplanckian gravitational scattering can be perturbatively calculated, black-hole formation occurs in the regime in which gravitational interactions are strong.
\end{enumerate}
Furthermore brane fluctuations (massless `branons') give rise to the same effect 1 (as in $\delta = 6$) and 2 (as in $\delta=4$)~\cite{crem}.
In its first stage with low statistics, LHC is particularly sensitive to the operator in eq.\eq{tau}, because its high dimensionality means
that the high energy of the LHC collisions is the key factor. 

In section~\ref{LHC} we show that the present low-statistics data about $pp\to jj$ already set a bound on the 
coefficient $8/M_{\cal T}^4$
of the effective operator\eq{tau}
which is significantly stronger than those obtained from any previous experiment, as summarized in table~\ref{tab:tau}.
In section~\ref{th} we discuss how $M_{\cal T}$ can be related to $M_D$ and $\delta$,
and derive explicit expressions for the full graviton-exchange amplitude,
including both gravitons at the ultraviolet end of the spectrum and gravitons that can be produced at LHC.
In section~\ref{full} we compare the full amplitude to LHC data.
Section~\ref{concl} contains our conclusions.

\section{Fit to the graviton-exchange effective operator}\label{LHC}
We compare the first LHC data to the new physics described by eq.s\eq{tau} and\eq{Upsilon}.
Since the $\delta$-dependent double trace term in ${\cal T}$ is irrelevant for collisions of particles with masses much smaller
than the LHC energy, our subsequent analysis applies to any number of extra dimensions (larger than 2) as well
as to branon effects.

\medskip

The tree-level  exchange of virtual gravitons described by the Lagrangian of eq.~(\ref{eq:tau}) mediates the processes
\beq  pp\to \ell^+\ell^-,\qquad pp\to \gamma\gamma, \qquad pp\to jj \,.\eeq
The experimental collaborations concentrated their sensitivity studies on the di-lepton and di-photon final states. 
However the corresponding cross sections are significantly lower than the $pp\to jj$ cross section, and this is the main factor that determines the observability of these signals
at the initial LHC stage with $\sqrt{s}=7\TeV$ and low luminosity. Indeed requiring final states with invariant mass greater than $1 \TeV$,  jets, leptons and photons with $\eta<2.5$, and additionally requiring $|\eta_{1}-\eta_{2}|<1.2$ for the jets, we find
\beq \sigma = \left(\frac{2\TeV}{M_T}\right)^8 \times\left\{\begin{array}{ll}
12.5\,{\rm pb} & \hbox{for $pp\to jj$}\\
10.4\,{\rm fb}  & \hbox{for $pp\to \mu^+\mu^-$}\\
21.3 \,{\rm fb} & \hbox{for $pp\to \gamma\gamma$}\\
\end{array}\right.\label{eq:sigma}~~.\eeq
This large difference in cross sections
is due partly to trivial flavor and color factors, and partly to the fact that the processes are mediated by the operator of dimension 8 in eq. (\ref{eq:tau}), which gives larger rates for the channels with more energetic initial states. In particular $pp\to jj$ benefits from the high energy of the initial partons $uu$ in the $t$-channel process.

In the following we shall show that, from the angular distribution of the jets, even with only 3.1 pb$^{-1}$ of integrated luminosity it was possible to obtain the dominant limit to date on the operator $\mathcal{T}$.

\smallskip

We study the effect of graviton-mediated amplitudes in the differential and in the total cross section, including interference effects between the SM and the new contributions. Both the total and the differential cross section are affected by NLO effects. However this sensitivity to higher order effects can be reduced by choosing a suitable kinematical quantity and restricting the analysis to certain kinematical regions.


\begin{figure}[t]
$$\includegraphics[width=0.45\textwidth]{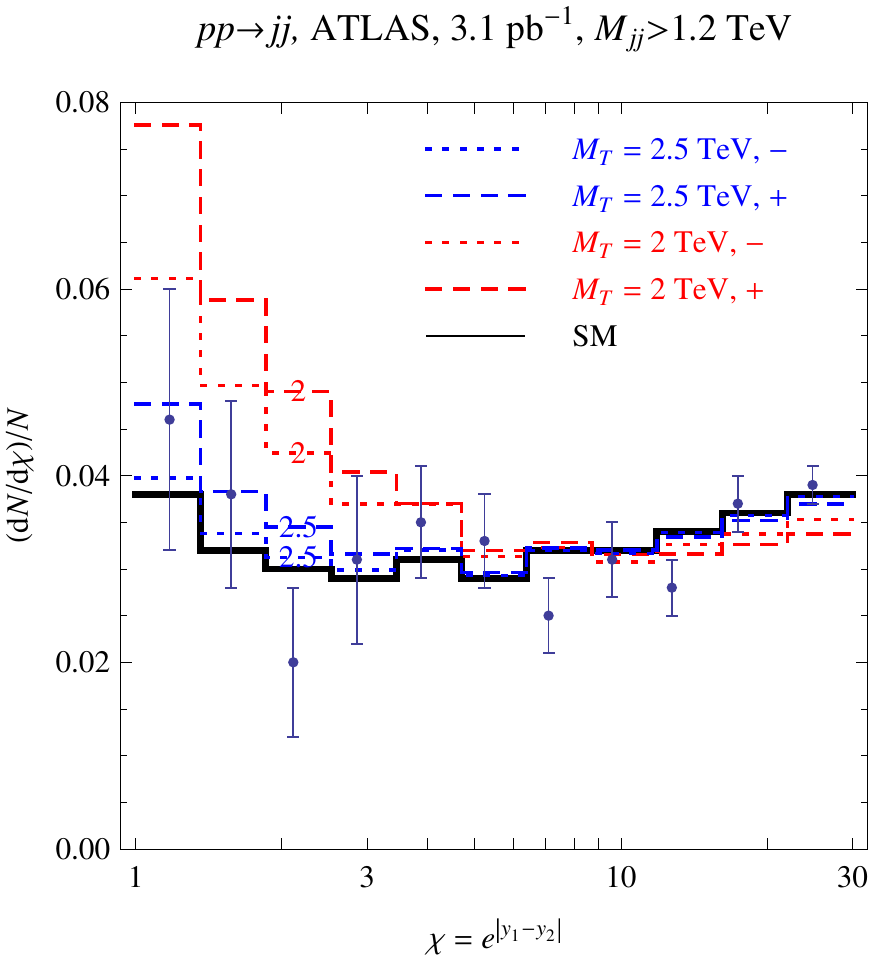}\qquad \includegraphics[width=0.465\textwidth]{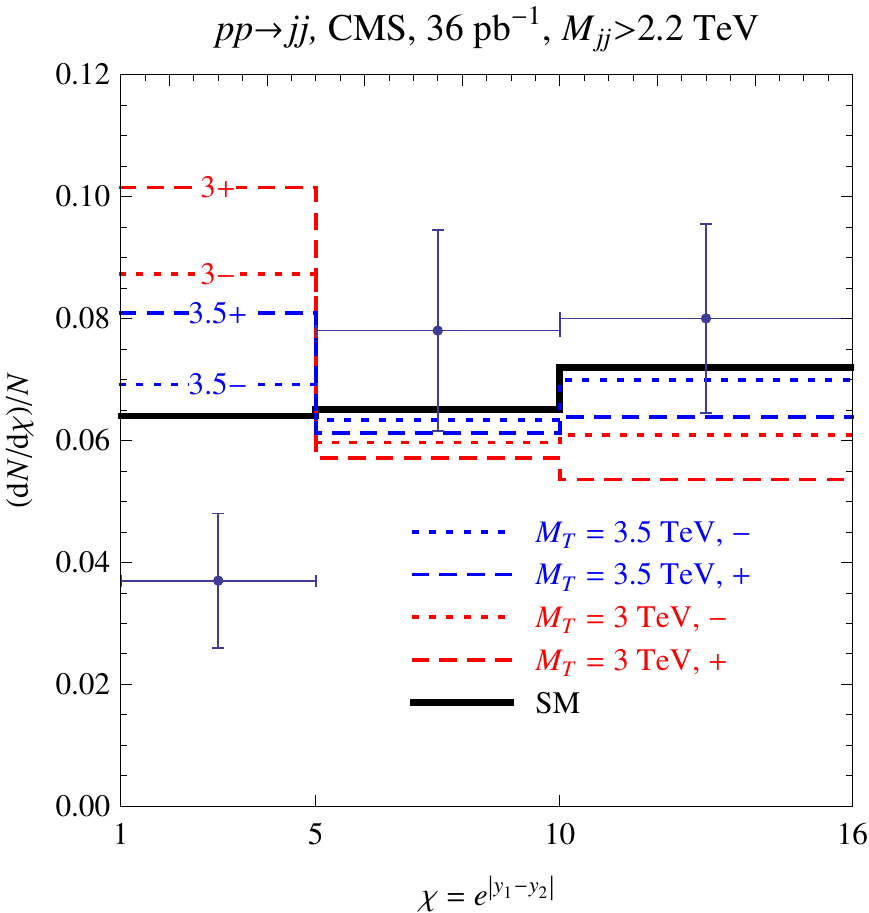}$$
\caption{\em Left (right): $pp\to jj$ angular distribution at ATLAS with $M_{jj}>1.2\TeV$
(at CMS with $M_{jj}>2.2\TeV$)
binned as a function of the angular distance
$\chi$.  The experimental data (crosses) are compared to the
SM prediction (black histogram) and to the
expectation including virtual graviton effects at tree level.
\label{fig:chi}}
\end{figure}


\begin{figure}[t]
$$\includegraphics[width=0.45\textwidth]{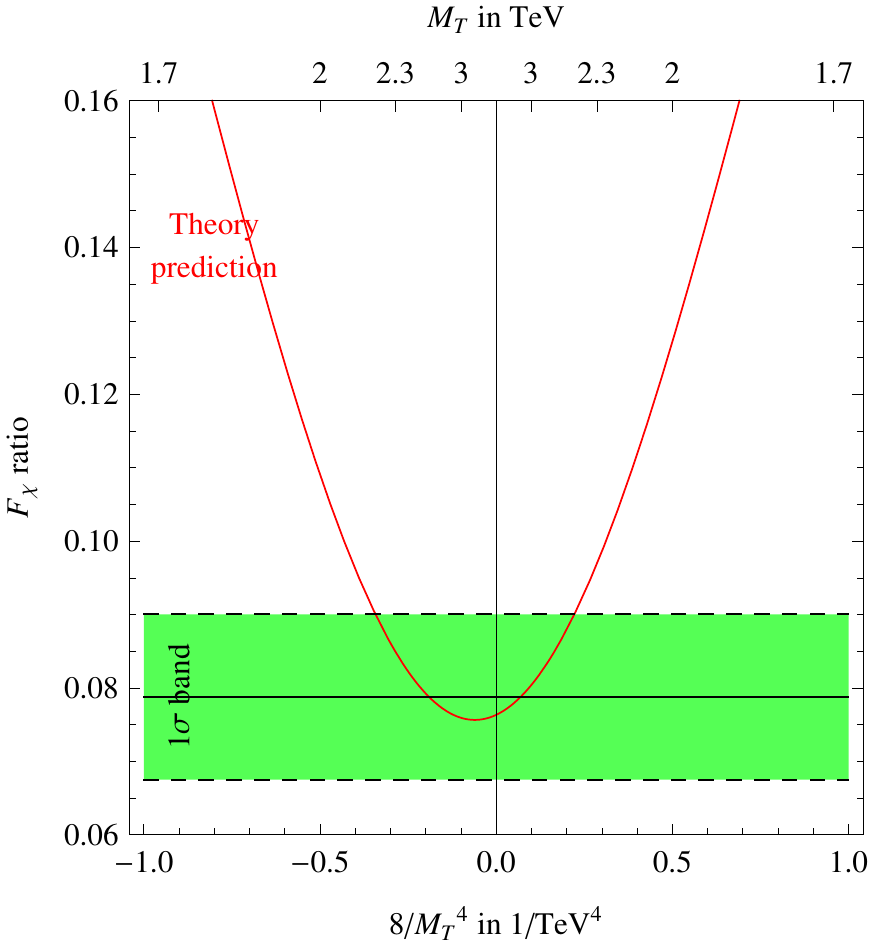}$$
\caption{\em Experimental values from ATLAS and theoretical values of the variable $F_\chi$ (fraction of $jj$ events with $M_{jj}>1.2\TeV$ in the central region).
\label{fig:Fchi}}
\end{figure}

ATLAS \cite{LHC7} and CMS~\cite{CMS,landsb} have searched for the effect of contact interactions in the angular  distribution of dijet events. Both collaborations have studied the centrality ratio distribution, and ATLAS also released the normalized distribution in several ranges of invariant mass of the jets  on the variable 
$$
\chi\equiv \exp|y_1 - y_2|\,,
$$
where $y_{1,2}$ are the two jet rapidities.  Due to the dominance of Coulomb-like scattering in the SM, these distributions are expected to be almost flat in the case of QCD, which helps to reduce the impact of smearing effects. Contact interactions, especially those in eq. (\ref{eq:tau}) being mediated by a spin-2 particle, have a different angular distribution with respect to QCD and result in a deviation from a flat distribution.



Data for the $\chi$ distribution from ATLAS are reported in fig.\fig{chi}a together with the SM expectation at next-to-leading order~\cite{LHC7}. 
Fig.\fig{chi}a shows also the effect of the graviton operator $\mathcal{T}$  for $M_T=2$ and 2.5 TeV for both positive or negative interference with the SM.

The prediction of the effect of the operator $\mathcal{T}$ has been obtained simulating the effect of this operator at the partonic level with {\sc MadGraph}~\cite{MadGraph} and CTEQ6L parton distribution functions.  We checked that showering and detector effects do not alter significantly the prediction. In particular we checked that with the current uncertainties on the data the limit on the contact interaction studied by ATLAS~\cite{LHC7} is reproduced at the partonic level within 20\%.

We compare data with the theoretical expectation and we compute the $95\%$ CL bound on the coefficient
of the ${\cal T}$ operator
by imposing
\beq \chi^2 = \sum_{i}^{\rm bins} \frac{(t_i(c_{\cal T}) - \mu_i)^2}{\sigma_{i~\rm stat}^2 +\sigma_{\rm syst}^2} < \chi^2_{\rm min} +3.84\,,\eeq
where $\mu_i$ are the experimental central values, $\sigma_{i~\rm stat}$ the statistical errors,
$\sigma_{\rm syst} \approx 0.003$ estimates the systematic uncertainties (we ignore possible correlations
between different bins) which are presently subdominant
and $t_i(c_{\cal T})$ are the theoretical predictions, computed for some values of $c_{\cal T}$ and fitted in each bin as a quadratic function of $c_{\cal T}=8/M_T^4$.  We find the bound $M_T>2.1\TeV$ reported in table~\ref{tab:tau}.
This significantly exceeds all previous bounds.

\smallskip

ATLAS~\cite{LHC7} reports also the observation on the quantity $F_\chi$, defined as the ratio between the events in the first four $\chi$ bins ($\chi < 3.3$)
with respect to the total 621 $jj$ events in the acceptance region.
The present experimental value is $F_\chi = 0.078\pm0.011$, to be compared to the SM prediction at NLO, $F_\chi^{\rm SM}= 0.076$ ~\cite{LHC7}.
We find that the variable $F_{\chi}$ captures well the effect of contact interactions, as it corresponds to comparing the cross-section in the central region for the SM and the contact interaction.  Indeed the bound on $M_T$ negligibly changes going from the full fit to the one-variable $F_\chi$ fit, as illustrated in fig.\fig{Fchi}.

The variable $F_{\chi}$ allows us to easily estimate how the sensitivity to
 $M_{T}$ improves with higher luminosity. Assuming that the measurement is dominated by the statistical error we find that with the luminosity of about $50$~pb$^{-1}$ currently collected by the LHC experiments the expected 95\% CL limit on $M_{T}$  is about 3 TeV.
 
 \medskip
 
CMS $pp\to jj$ data after 36 pb$^{-1}$ have been recently presented~\cite{landsb} and are here plotted in fig.\fig{chi}b.
We can reliably estimate the resulting bound, $M_T > 3.4\TeV$, as reported in table~\ref{tab:tau}.
This is comparable to the sensitivity, $3.2\TeV$, despite the
apparent mild statistical fluctuation in the first bin.
From ATLAS $pp\to jj$ data after 36 pb$^{-1}$ \cite{ATLASjetjet36} we estimate $M_T >3.2\TeV$.

\medskip

We can compare the sensitivity of the dijet channel to those of the
$ pp\to \ell^+\ell^-$ and $\, pp\to \gamma\gamma$ channels
considered by the experimental collaborations. 
CMS~\cite{CMSgammagamma} finds $M_T>1.8\TeV$ from $pp\to\gamma\gamma$ after 36/pb of integrated luminosity.
Ref.~\cite{CMSgammagamma}   reports a 95\% C.L. sensitivity in the $\gamma\gamma$ channel to $M_{T}\simeq 3 \TeV$ for more than 150 pb$^{-1}$ at 10 TeV center of mass energy and~\cite{CMSmumu} claims a sensitivity of the leptonic channel to $M_{T}\simeq 3 \TeV$ with 100 pb$^{-1}$ of 14 TeV data. 
The proposed measurements essentially consist in counting events with large invariant mass, as in eq.\eq{sigma}.
The $pp\to jj$ signal already reached the same sensitivity with current  center of mass energy of 7 TeV and current 36 pb$^{-1}$ of integrated luminosity, and will remain the most sensitive channel until systematic uncertainties will dominate the error on the measured angular distribution.

\medskip

Finally, we computed the bound on the dimension-6 operator $\Upsilon$ of eq.\eq{Upsilon}
generated by graviton exchange at loop level. 
The result is shown in table~\ref{tab:dim6} together the other existing bounds. With the published data we find a bound from dijets  at LHC
that is comparable to the bound from Tevatron and strongly
subdominant with respect to the bound from LEP. 
Even with the data  with 36 pb$^{-1}$ of integrated luminosity
we get a bound subdominant with respect to LEP, although significantly larger that existing limits from Tevatron.

\begin{figure}[t]
$$\includegraphics[width=\textwidth]{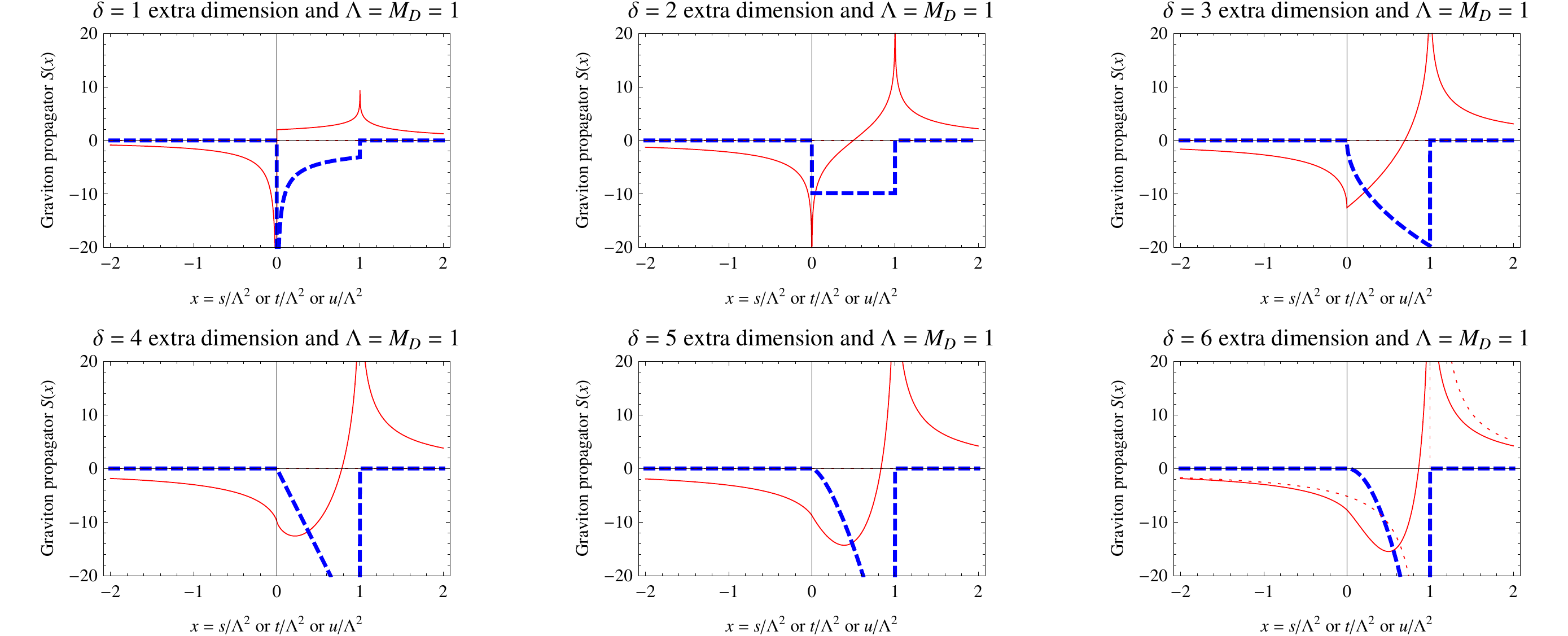}$$
\caption{\em Real part (solid red curve) and imaginary part (dashed blue curve) of ${\cal S}(x)$ in units $\Lambda =M_D =1$.
The dotted line in the $\delta=6$ panel shows the single-pole approximation
(one graviton with mass $\Lambda$) that holds in the limit $\delta \to\infty$.
\label{fig:Ss}}
\end{figure}

\section{Tree-level graviton exchange}\label{th}
In view of its experimental significance, we reconsider the theory behind eq.\eq{tau} and the approximation of
tree-level graviton exchange with an effective operator.
In full generality, tree-level graviton-exchange (fig.\fig{treeloop}a) leads to a scattering amplitude of the form
\beq\label{eq:T}
{\cal A} = {\cal S}(s) \left( T_{\mu\nu}T^{\mu\nu}-
\frac{T^\mu_\mu T^\nu_\nu}{\delta +2}\right) .
\eeq
The function ${\cal S}$ is obtained by summing over all the Kaluza-Klein (KK) tower of gravitons.
As will be discussed below, if the typical energy resolution of the experiment is broader than the mass separation between two KK states, the sum can be approximated as an integral over the extra-dimensional momentum $q$ of the graviton.
Such integral is UV divergent for $\delta>1$ extra dimensions. 
So we regularize the integral by including only KK excitations with mass $m = |q|$ below an arbitrary cut-off $\Lambda$, which parametrizes the onset of the unknown
quantum-gravity physics.  A small (large) ratio $\Lambda/M_D$ effectively means that quantum gravity is weakly (strongly) coupled~\cite{stru}.
The use of the cutoff allows for a comparison of the experimental limits on the operator\eq{tau} with the searches for real graviton emission in missing $p_T$ events. Cutting off the integral, we find
%
\beq
{\cal S}(s)=\frac{1}{M _D^{2+\delta}}
\int_{|q|<\Lambda}\frac{d^\delta q}{s-q^2+i\varepsilon}=
\frac{\pi^{\delta/2}\, \Lambda^{\delta-2}\, }{\Gamma(\delta/2) \, M_D^{2+\delta}}F_\delta({s \over \Lambda^2})
\label{eq:estau}
\eeq
where $\Gamma$ is the Euler function and $F_\delta$ is recursively defined as
\beq
F_{\delta +2}(x) =  xF_\delta(x) -\frac{2}{\delta}\eeq
and\footnote{Equivalent, but more explicit, expressions for $F_{1,2}$ are
$$\Im F_1= -\pi/\sqrt{x},\qquad \Im F_2 = -\pi \quad\hbox{for $0<x<1$ and zero otherwise}$$
$${\rm Re}\, F_1 = \left\{\begin{array}{ll}
 \frac{1}{\sqrt{x}} \ln \left| \frac{\sqrt{x} +1}{\sqrt{x} -1}\right|  & {\rm for}~x>0\cr
 \frac{1}{\sqrt{-x}} \left[ 2{\rm arctan} \left( \sqrt{-x}\right) -\pi \right] & {\rm for}~x<0
 \end{array}\right.\qquad
 {\rm Re}\, F_2 = -\ln \left| 1-\frac{1}{x} \right| .$$}

\beq F_1(x)= \frac{2}{\sqrt{x}}\, {\rm arctanh} \frac{1}{\sqrt{x}} ,\qquad F_2(x) = -\log \left( 1-\frac{1}{x}\right) .\eeq
%
%
Fig.\fig{Ss} shows the behavior of the real (solid line) and imaginary (dashed line) parts of ${\cal S}$, for various values of $\delta$. In the case of $t$-channel exchange, the variable of the function ${\cal S}$ is negative and no imaginary part is developed, since the exchanged graviton cannot be on-shell. 
For $\delta >2$ the integral is dominated by the heaviest graviton with mass $m\approx \Lambda$ and thus, for $s\ll \Lambda^2 $, the function ${\cal S}$ can be treated as a constant with no momentum dependence and the scattering amplitude can be approximated by the effective operator ${\cal T}$ of eq.\eq{tau} with a coefficient which is usually defined as~\cite{hewett}
\beq
{\cal
S}(s\ll \Lambda^2) =
\left\{\begin{array}{ll }\displaystyle
 \frac{\pi^{\delta /2}}{(1-\delta/2)\Gamma (\delta /2)} \;
               \frac{\Lambda^{\delta -2}}{M_D^{\delta +2}}   \equiv \frac{8}{M_{\cal T}^4}   &  {\rm for}~\delta >2  \\
               \displaystyle
\frac{\pi}{M_D^4}\ln \frac{s}{\Lambda^2}      &   {\rm for}~\delta =2\\
               \displaystyle       \frac{-i\pi}{M_D^3\sqrt{s}}  &  {\rm for}~\delta=1  \\
               \end{array}\right.
\label{eq:op_lhc}
\eeq
However, in view of the high dimensionality of the operator, the dominant LHC bound comes from the highest energy events,
and it is appropriate to retain the full amplitude, including the dependence on the cut-off $\Lambda$.

We would like now to comment on the validity of approximating the sum over virtual gravitons with an integral. 

\subsection{$\delta=1$}

It is well known that gravity at macroscopic scales and astrophysical considerations strongly constrain the cases $\delta =$ 1, 2, and 3. The corresponding fundamental mass $M_D$ can lie around the weak scale only if the theory is modified in the infrared. This can be achieved by introducing a warping factor~\cite{rs1} with a small mass parameter $\mu$ (of a few MeV) which lifts the lightest KK mode of the graviton
(and characterizes the KK graviton mass splitting, since $m_n\simeq \pi n\mu$ for $n\gg 1$), without modifying the UV behavior of the theory and its collider predictions~\cite{Plehn}. 

Let us first consider the case $\delta =1$, in which the KK summation can be explicitly performed with the result~\cite{kiss}
\beq
{\cal S}(s) =  \frac{1}{\Lambda_\pi^2} \sum_n \frac{1}{s-m_n^2+i m_n\Gamma_G (m_n)}
=- \frac{\pi }{M_5^3\sqrt{s}}K 
\label{eq:discrete}
\eeq
\beq
 K = \frac{\sin2A+i\sinh 2\epsilon}{2(\cos^2A+\sinh^2\epsilon )}\qquad A=\pi \left( \frac{\sqrt{s}}{\Delta m} +\frac{1}{4}\right) \qquad \epsilon = \left. \frac{\pi \Gamma_G}{2\Delta m}\right|_{m=\sqrt{s}}.
 \label{eq:kisdef}
\eeq
Here $\Lambda_\pi$ is the interaction scale of individual gravitons, related to the fundamental mass of the 5-dimensional theory $M_5$ by~\cite{Plehn}
\beq
\Lambda_\pi^2 =\frac{{M_5}^3}{2\pi \mu}.
\eeq
The mass splitting between KK gravitons $\Delta m$ and the decay width of the $n$-th KK graviton $\Gamma_G (m_n)$ are given by
\beq
\Delta m = \pi \mu  \qquad \Gamma_G (m_n) =  \frac{c\, m_n^3}{\pi \Lambda_\pi^2},
\eeq
where $c=1/80,~1/320$, and $1/960$ for graviton decays into a massless vector, Weyl fermion, and conformally-coupled
real scalar, respectively~\cite{han}.  Consequently, we find
$c = 283/960$ after summing over all SM particles.
The parameter $\epsilon$ in eq.\eq{kisdef}, which measures the relative separation of the individual graviton resonances ($\epsilon \ll 1$ means well separated resonances, $\epsilon\circa{>}1$ means overlapping resonances), is given by
\beq
\epsilon = c \left( \frac{\sqrt{s}}{M_5}\right)^3.
\eeq
Therefore $\epsilon$ remains finite in the limit $\mu \to 0$, which corresponds to sending the compactification volume to infinity ($M_{\rm Pl} \to \infty$).

The expression of ${\cal S}$ in eq.\eq{discrete} is a rapidly oscillating function. However, we are interested in the case in which the energy spread of the initial and final states is broader than the mass separation $\mu$. It is then convenient to average eq.\eq{discrete} within one oscillation period, obtaining the smoothly varying function\footnote{We use $$\frac{1}{2\pi}\int_0^{2\pi} dx\, \frac{\sin 2x +a}{\cos^2x+b}=\frac{a}{\sqrt{b(1+b)}}.$$}~\cite{Plehn}
\beq
\langle {\cal S}\rangle = - \frac{i\pi}{M_5^3\sqrt{s}}.
\label{eq:risdis}
\eeq

We can now take an alternative approach and work directly in the continuum, by replacing the discrete KK summation with an integral\footnote{\label{due}We use$$\lim_{\epsilon \to 0} \frac{1}{x+i\epsilon}= P\left(\frac{1}{x}\right) - i\pi \delta(x).$$}
\beq
{\cal S}(s) =  \frac{1}{\Lambda_\pi^2} \int \frac{dm}{\pi \mu}~ \frac{1}{s-m^2+i m\Gamma_G (m)} \stackrel{\Gamma_G \to 0}{\simeq} - \frac{i\pi}{M_5^3\sqrt{s}}.
\label{eq:riscon}
\eeq
Therefore, the procedure of integrating in the continuum, eq.\eq{riscon}, gives exactly the same result as the averaged summation in eq.\eq{risdis}.
This shows that, as long as the energy resolution is broader than the mass separation, it is perfectly adequate to treat virtual gravitons as a continuum.

Let us now consider the modulus square of the expression in eq.\eq{discrete}, averaged over an oscillation period\footnote{We use $$\frac{1}{2\pi}\int_0^{2\pi} dx\, \frac{\sin^2 2x +a}{(\cos^2x+b)^2}=\left[ 2+ \frac{a}{4b(1+b)}\right] \frac{1+2b}{\sqrt{b(1+b)}}-4.$$}
\beq
\langle |{\cal S}|^2\rangle = \frac{\pi^2}{M_5^6 s} \left( 1+\frac{4}{e^{4\epsilon}-1}\right) \stackrel{\epsilon \to 0}{\simeq}\frac{1}{\epsilon} \left( \frac{\pi}{M_5^3 \sqrt{s}}\right)^2.
\label{eq:uffa}
\eeq
While for $\epsilon >1$ we find $\langle |{\cal S}|^2\rangle \simeq |\langle {\cal S}\rangle |^2$, in the relevant case of small $\epsilon$ we obtain that eq.\eq{uffa} leads to an enhancement of a factor $1/\epsilon$. Note that the enhanced term in eq.\eq{uffa} has a lower order in powers of graviton coupling constants than expected for a scattering process, because it corresponds to the production of real gravitons. 

The same result can be obtained also by calculating $|{\cal S}|^2$ in the continuum. If we are interested in the real production of well-separated narrow resonances, we can neglect interference effects. Then the calculation in the continuum, for $\epsilon <1$, gives\footnote{We use
$$\lim_{\epsilon \to 0} \frac{\epsilon}{x^2+\epsilon^2}= \pi \delta(x).$$}
\beq
|{\cal S}|^2 =\frac{1}{\Lambda_\pi^4} \int \frac{dm}{\pi \mu} ~ \frac{1}{(s-m^2)^2+m^2\Gamma_G^2}  \stackrel{\Gamma_G \to 0}{\simeq}\frac{1}{\epsilon} \left(\frac{\pi}{M_5^3 \sqrt{s}}\right)^2 .
\eeq
The result of the calculation in the continuum agrees with the discrete summation in eq.\eq{uffa}, when initial and final particle states are spread in energy more than the KK mass separation.

\medskip

As mentioned above,  the Feynman diagram in fig.\fig{treeloop}a includes two effects:
a) $2\to 2$ scattering processes mediated by virtual gravitons, and b) $2\to 1\to 2$ production of one 
graviton KK resonance with mass equal to $\sqrt{s}$ that eventually decays into SM particles.
The enhancement in eq.\eq{uffa} is the contribution from process b).
In the $\delta=1$ scenario we are considering, the graviton decays well inside the detector,
such that process b) must be included and there are no missing-energy signals
(a point missed in previous works on the topic).

On the contrary, in the $\delta>1$ scenarios considered in the next section, 
KK gravitons typically decay far away from the detectors, such that process b) does not contribute to
$2\to 2$ scatterings observed at LHC.


\medskip

%

\subsection{$\delta>1$}
The previous result can be generalized to $\delta >1$. The amplitude smoothed over scattering wave packets broader than the mass splitting between KK gravitons is obtained by replacing the discrete summation with an integral
\beq
{\cal S}(s) =\frac{1}{M_{\rm Pl}^2}\sum_i \frac{1}{s-m_i^2+im_i\Gamma_G(m_i)} \to
\frac{2\pi^{\delta /2}}{\Gamma (\delta /2)M _D^{2+\delta}}
\int_0^\Lambda dm\, \frac{m^{\delta -1}}{s-m^2+im\Gamma_G(m)},
\eeq
where $M_{\rm Pl}$ is the reduced Planck mass.
Writing the graviton propagator in the narrow-width approximation and using the relation in footnote~\ref{due}, we obtain an expression for $\langle {\cal S} \rangle$ that is identical to eq.\eq{estau}.

For generic $\delta$, the graviton width is $\Gamma_G (m) =c\, m^3/\pi M_{\rm Pl}^2$  and the mass difference is 
\beq
\Delta m = \frac{\Gamma (\delta /2) M_D^{2+\delta}}{2\pi^{\delta /2}M_{\rm Pl}^2m^{\delta -1}}.
\eeq 
Here we are considering the case in which the KK graviton spectrum is not distorted in the infrared ($\mu =0$). Analogously to the $\delta =1$ case, we can define
\beq
\epsilon \equiv \left. \frac{\pi \Gamma_G}{2\Delta m}\right|_{m=\sqrt{s}} = \frac{\pi^{\delta/2}c}{\Gamma (\delta /2)} \left( \frac{\sqrt{s}}{M_D}\right)^{2+\delta}.
\eeq
Note that $\epsilon <1$ as long as the low-energy effective theory can be trusted ($\sqrt{s}<M_D$), showing that the graviton resonances are narrow and well separated.  Using the narrow-width approximation (see relation in footnote 4) we find that the leading contribution is
\beq
\langle |{\cal S}(s)|^2 \rangle =  \frac{\left( {\rm Im}\, {\cal S} \right)^2}{\epsilon} .
\label{eq:estau3}
\eeq
As before, this term has to be interpreted as the production of a graviton with mass $\sqrt{s}$. Since the graviton decays well beyond the detector, this term contributes to ``missing energy" and not to the signal we are considering and should be subtracted from the final result. Effectively, the rate of interest is obtained by taking the modulus square of eq.\eq{estau}.
The situation can be different in intermediate scenarios with $\mu>0$ and shorter graviton life-time;
a life-time comparable to the detector size would lead to $2\to 2$ signals with displaced-vertex.

\begin{figure}[t]
$$\includegraphics[width=0.45\textwidth]{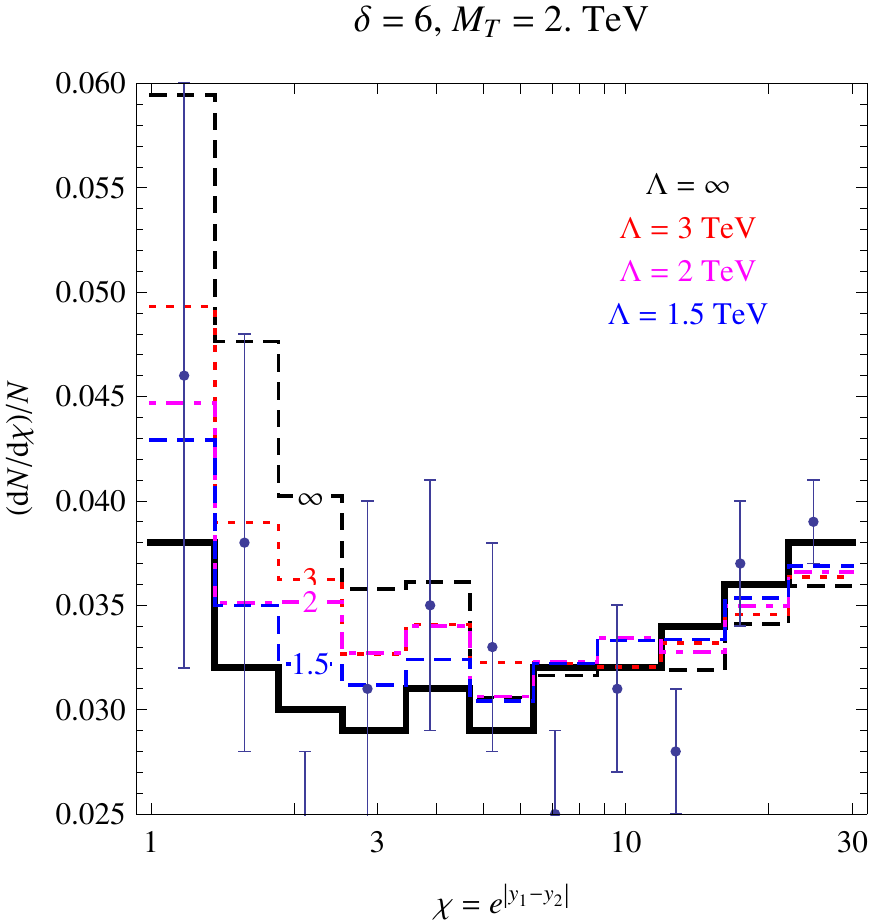}\qquad \includegraphics[width=0.45\textwidth]{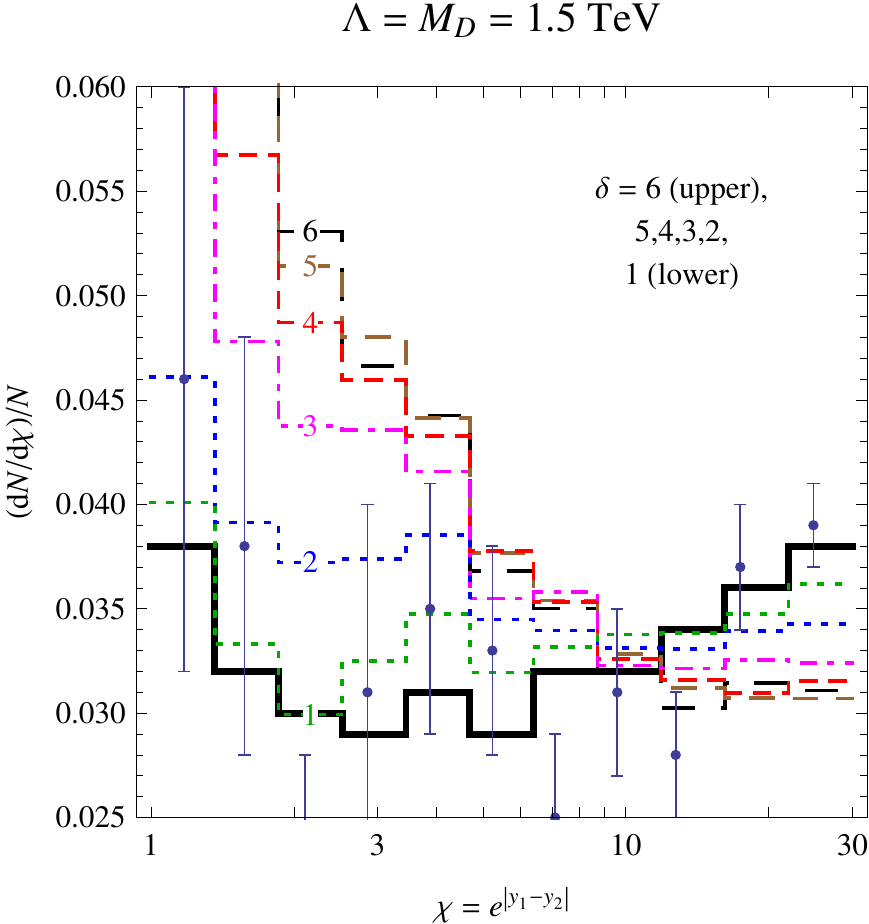}$$
\caption{\em Left: $pp\to jj$ angular distribution for fixed $\delta=6$, $M_T=2\TeV$, $M_{jj}>1.2\TeV$
 and different values of $\Lambda$
(as indicated) and consequently of $M_D$. The effective-operator ${\cal T}$ is formally reproduced in the limit $\Lambda\to\infty$.
Right: dependence on the number $\delta$ of extra dimensions at fixed $M_D=\Lambda=1.5\TeV$.
The data are from ATLAS \cite{LHC7}.
\label{fig:chifull}}
\end{figure}

\begin{figure}[p]
$$\includegraphics[width=0.45\textwidth,height=0.38\textwidth]{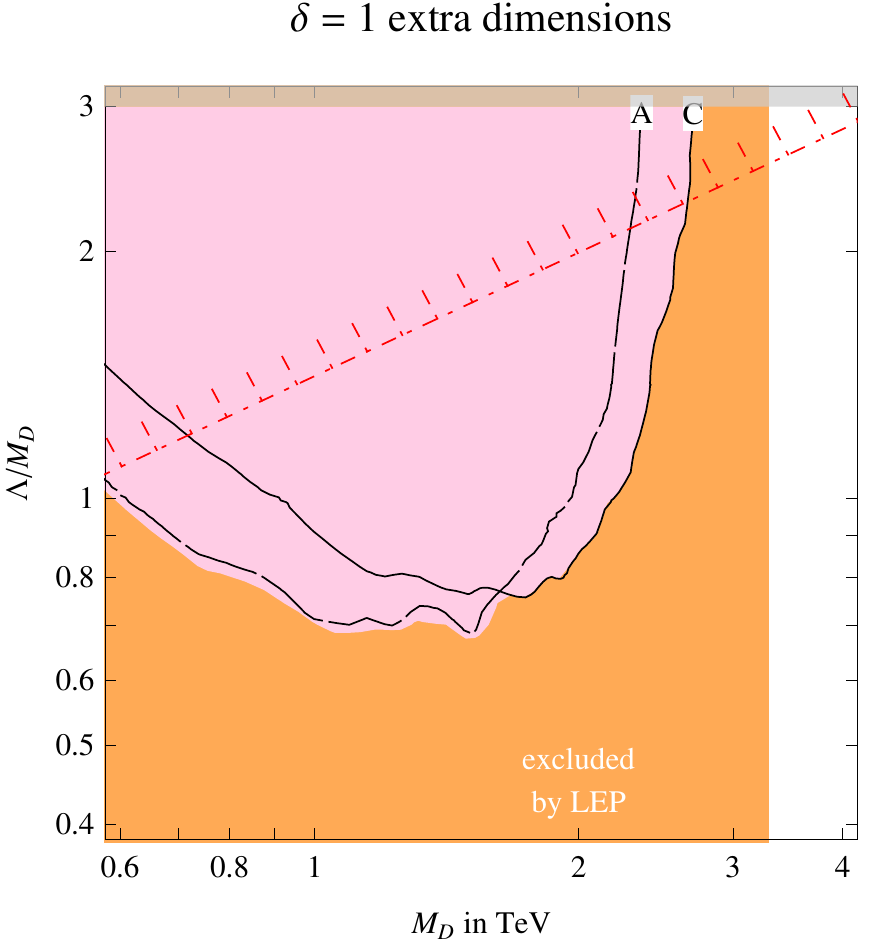}\qquad
\includegraphics[width=0.45\textwidth,height=0.38\textwidth]{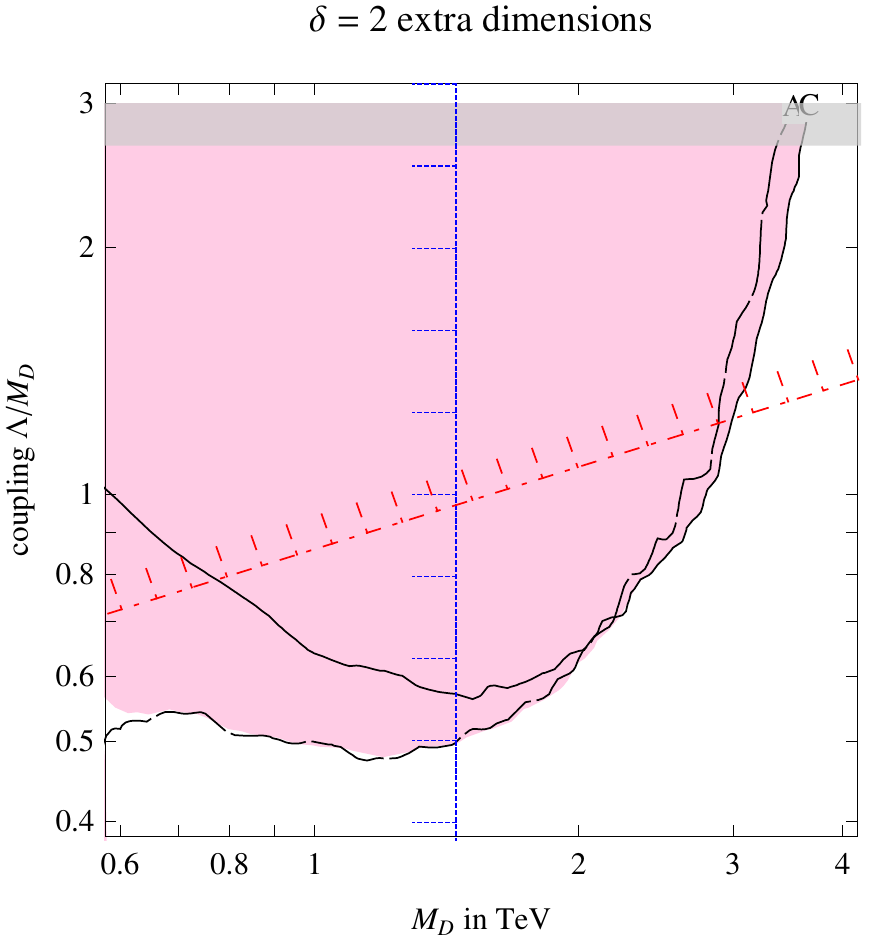}$$
$$\includegraphics[width=0.45\textwidth,height=0.38\textwidth]{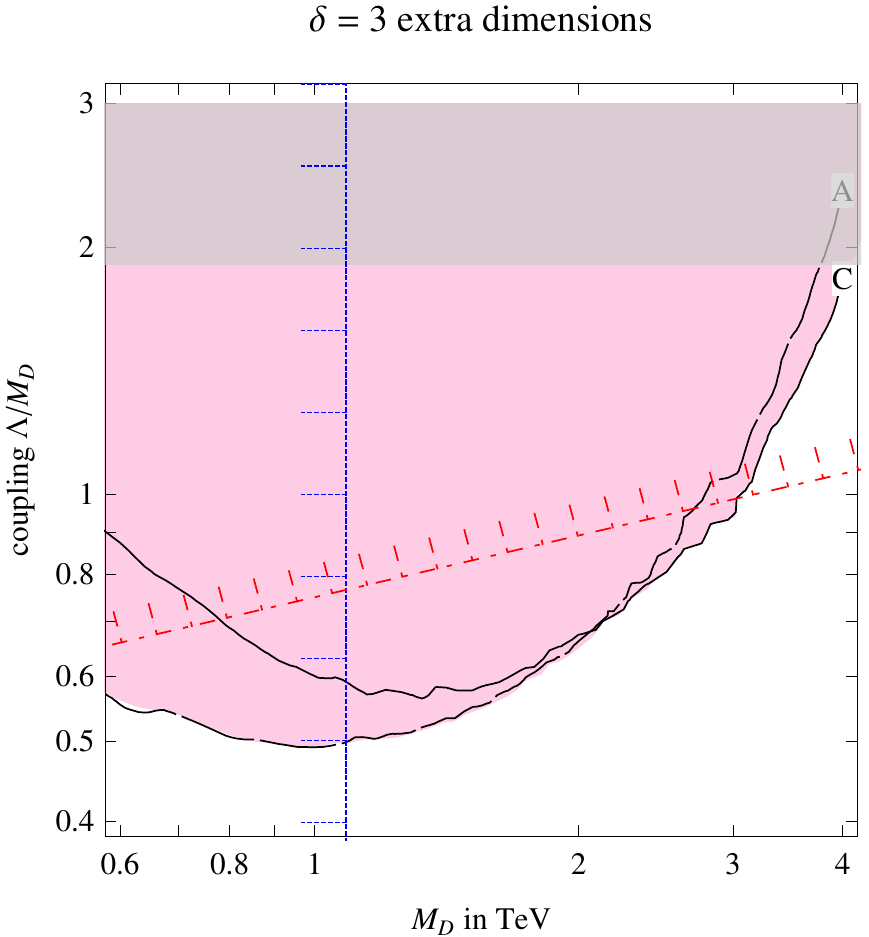}\qquad
\includegraphics[width=0.45\textwidth,height=0.38\textwidth]{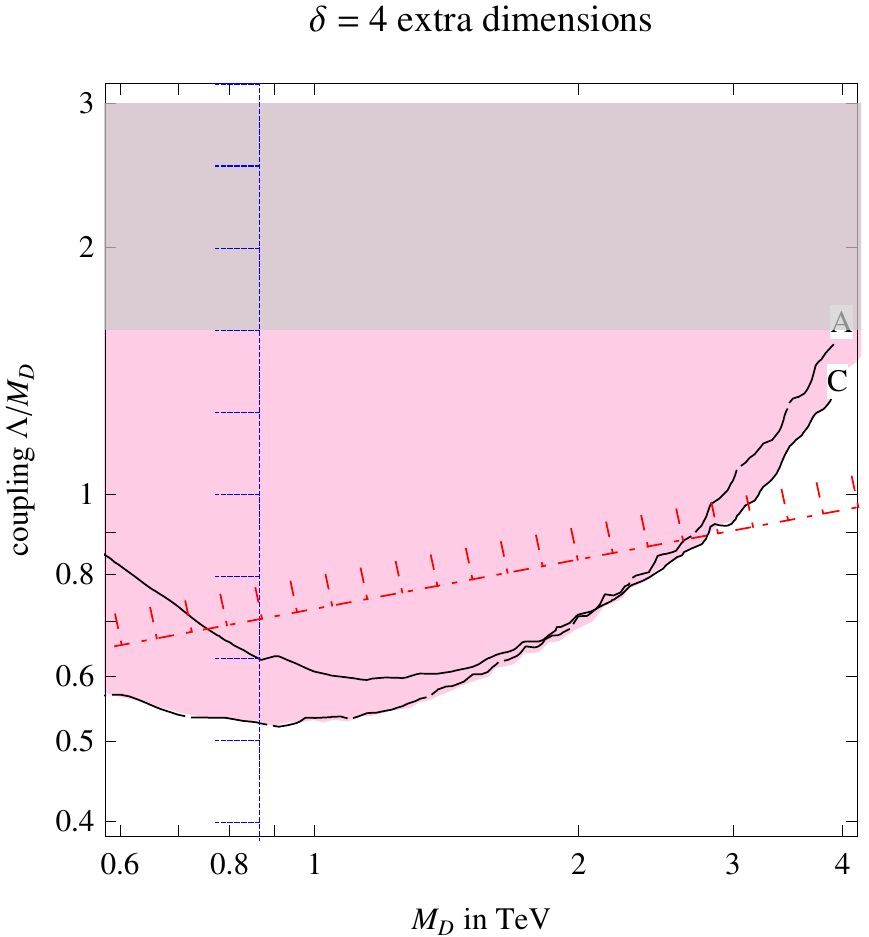}$$
$$\includegraphics[width=0.45\textwidth,height=0.38\textwidth]{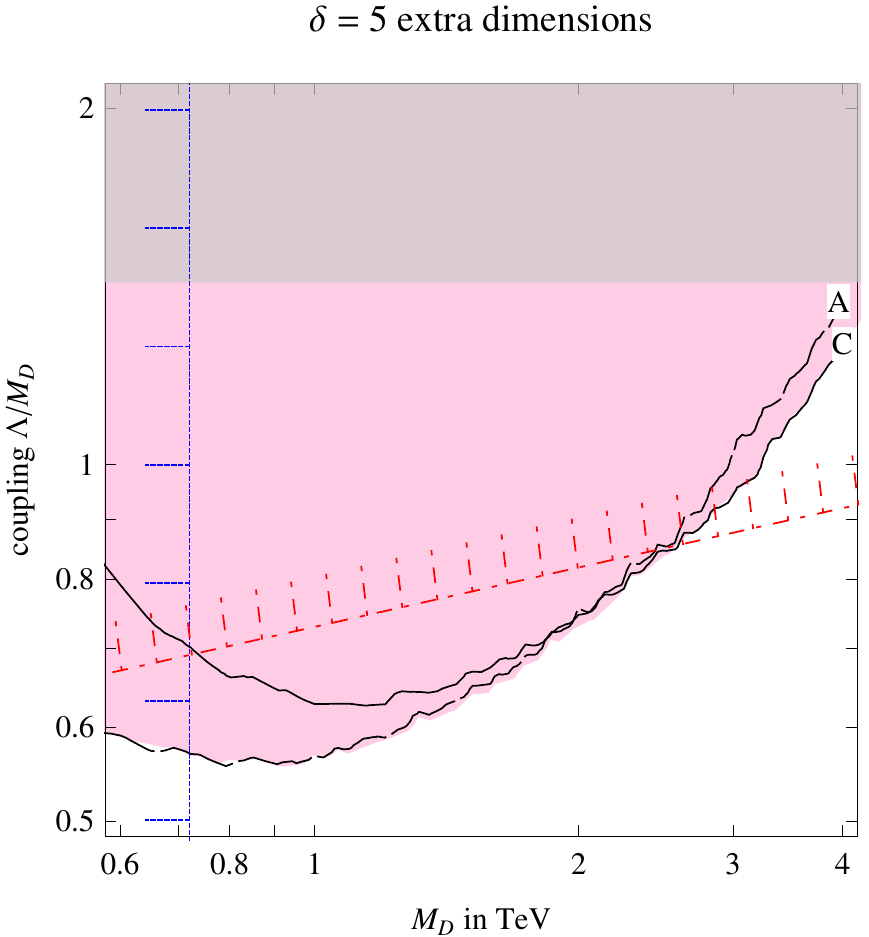}\qquad
\includegraphics[width=0.45\textwidth,height=0.38\textwidth]{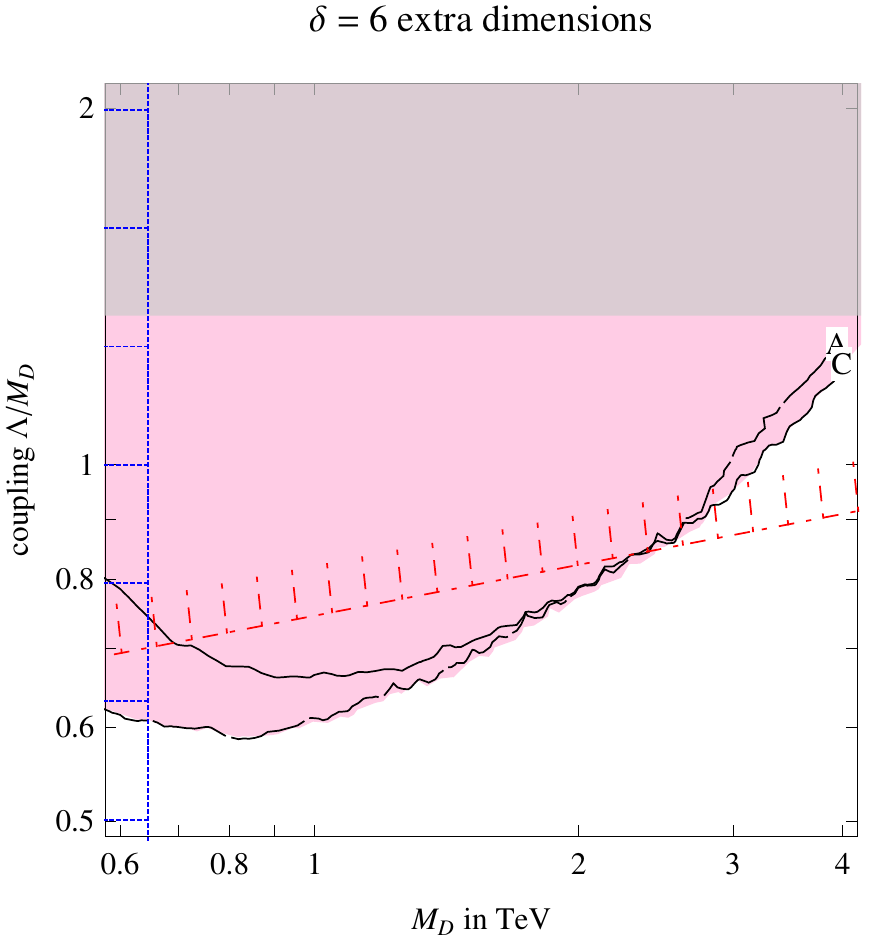}$$
\caption{\em
The shaded area is the bound from virtual graviton exchange at 
CMS (continuous line denoted as `C', data after {\rm 36/pb}),
ATLAS (long-dashed line denoted as `A', data after {\rm 36/pb}). 
Vertical blue line: bound from graviton emission (as summarized in table~1 of~\cite{stru}).
Red line: Naive Dimensional Analysis estimate of LEP bound from loop graviton exchange.
Upper shading: NDA estimate of the non-perturbative region.
\label{fig:bounds}}
\end{figure}

\section{Fit to the full graviton-exchange amplitude}\label{full}
Formul\ae{} for the cross sections from tree-level graviton effects in any number of extra dimensions can be found in the appendix of ref.~\cite{Plehn}. We implement them in {\sc Pythia}8~\cite{pythia} and verify that in the effective-operator
approximation (${\cal S} = 8/M_T^4$) the various distributions reproduce the ones previously obtained with {\sc MadGraph}
and that hadronization and jet reconstruction negligibly affect the observables we consider.

We can now compare the data with the full graviton-exchange amplitude, computed in terms of the
cut-off $\Lambda$, defined to be the maximal KK graviton mass.
Even for $\delta=1$ the correct treatment of $|{\cal S}|^2$ in the $s$-channel
 is numerically irrelevant for this work,
where we consider the $pp\to jj$ signal which is dominated by the $uu$ initial state
which has no $s$-channel.

\medskip

Fig.\fig{chifull}a shows how the theoretical prediction changes with the cut-off $\Lambda$ keeping fixed the
coefficient of the effective operator ${\cal T}$ to be $M_T=2\TeV$, around the present bound:
 the full amplitude must be used unless $\Lambda\gg M_D$.
 Fig.\fig{chifull}b shows how the theoretical prediction changes with $\delta$ keeping fixed $M_D$ and $\Lambda$.

\medskip

The results of our fit are shown in fig.\fig{bounds}, as functions of $M_D$ and of the ratio $\Lambda/M_D$.
The 95\% CL bound is defined as $\chi^2 < \chi^2_{\rm SM}+3.84$.
As previously discussed, the ratio $\Lambda/M_D$ effectively parameterizes the unknown strength of the
full quantum-gravity theory. The gray area at larger $\Lambda/M_D$ covers the region estimated to be
non-perturbative according to naive dimensional analysis~\cite{stru}.

The shaded area covers the region excluded by the angular distribution at $M_{jj}>1.2\TeV$.
The dashed curve also shows the exclusion obtained considering only the $F_\chi$ ratio:
it gives a good approximation to the full fit in the region with larger $\Lambda$ where the effective operator
approximation is valid; but fails in the region with lower $\Lambda$.

For comparison, the other two lines show:
\begin{itemize} 
\item the combined Tevatron-LEP bound from graviton emission (vertical blue lines;
computed ignoring the dependence on $\Lambda$).

\item the LEP bound on loop graviton exchange
(red line), estimated according to naive dimensional analysis.
\end{itemize}
For $\delta=1$ the LHC bound $M_D\circa{>}1.5\TeV$ remains subdominant
with respect to the bound from
$e^-e^+\to f\bar f$ scatterings at LEP2, that we estimate to be 
$M_D\circa{>}3.4\TeV$.
For $\delta >1$ already the first ATLAS data at 3.1 pb$^-1$ explore new regions of the parameters space of gravity in extra dimensions.
The new data at 36 pb$^-1$ provide stronger bounds.

\section{Conclusion}\label{concl}
We found that the very first LHC data about
$pp\to jj$, despite the low statistics and the uncertainties intrinsic in the hadronic nature of the final state, improve significantly
previous bounds on the coefficient of the effective dimension-8 operator ${\cal T}$ generated by virtual graviton or branon exchange,
and predicted by theories with extra dimensions.  This arises thanks to the high dimensionality of the operator,
which rewards the higher energy of LHC with respect to previous colliders.

In a second part of the work we went beyond the effective-operator approximation
and computed the  full amplitude generated by tree level graviton exchange
in terms of a cut-off parameter $\Lambda$, which is the maximal KK graviton mass.  
We clarified that the enhanced effect of lighter gravitons that can be produced on-shell must 
be included only when such gravitons decay within the detector.
Fig.\fig{bounds} shows the resulting LHC bounds in the $(M_D, \Lambda/M_D)$ plane.

\bigskip

\paragraph{Acknowledgements} We thank Riccardo Rattazzi for discussions about the physics of extra-dimensions and for his suggestions. We are grateful to Georgios Choudalakis and Frederik Ruehr for discussions about the published results of ATLAS, and to Greg Landsberg about CMS data presented in~\cite{landsb}. RF is also grateful to Diego Blas and Michele Redi for many discussions on gravity and extra-dimensions.
This work was supported by the ESF grant MTT8 and by SF0690030s09 project.
The work of RF is supported by the Swiss National Science Foundation under contract No. 200021-116372.
The work of PL is supported in part by the European Programme 'Unification in the LHC Era', contract PITN-GA-2009-237920 (UNILHC).
\small
\begin{multicols}{2}

\end{multicols}
\end{document}